\documentclass[a4paper,11pt]{article}
\usepackage{pos}
\usepackage{subcaption}
\title{Enhancing CTAO Monitoring and Alarm Subsystems in Distributed Environments Using ServiMon}
\ShortTitle{Enhancing CTAO Monitoring and Alarm Subsystems Using ServiMon}

\manuallySeparateAuthors
\author*[a]{Kevin Munari}
\author[a]{, Alessandro Costa}
\author[a]{, Federico Incardona}
\author[a]{, Emilio Mastriani}
\author[a]{, Sebastiano Spinello}
\author[b]{, Stefano Germani}
\author[a]{, and Pietro Bruno}
\author[c]{, for the CTAO Consortium}

\affiliation[a]{INAF, Osservatorio Astrofisico di Catania, Via S. Sofia 78, I-95123 Catania, Italy}
\affiliation[b]{Universit\`a di Perugia, Dipartimento di Fisica e Geologia, Italy}
\affiliation[c]{See \url{www.ctao.org}}

\emailAdd{kevin.munari@inaf.it}
\emailAdd{alessandro.costa@inaf.it}

\abstract{ServiMon is a scalable data collection and auditing pipeline designed for service-oriented, cost-efficient quality control in distributed environments, including the CTAO monitoring, logging, and alarm subsystems. Developed within a Docker-based architecture, it leverages cloud-native technologies and distributed computing principles to enhance system observability and reliability.

At its core, ServiMon integrates key technologies such as Prometheus, Grafana, Kafka, and Cassandra. Prometheus serves as the primary engine for real-time performance metric collection, enabling efficient monitoring across multiple nodes. Grafana provides interactive, service-oriented data visualization, facilitating system performance analysis. Additionally, Kafka and Cassandra expose system metrics via the JMX Exporter, offering critical insights into infrastructure availability and performance.

This contribution exposes how ServiMon could provide an enhancement on scalability, security, and efficiency in a distributed computing environment, such as the CTAO monitoring, logging, and alarm subsystems. This integrated approach not only ensures robust real-time monitoring, but also optimizes operational costs. Furthermore, ServiMon’s ability to generate large volumes of diverse data over time provides a strong foundation for predictive maintenance. By incorporating stochastic and approximate computing techniques, it enables proactive failure detection and system optimization, minimizing downtime and maximizing telescope availability.}

\ConferenceLogo{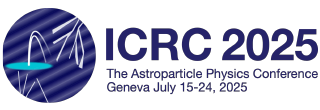}

\FullConference{39th International Cosmic Ray Conference (ICRC2025)\\
 15–24 July 2025\\
Geneva, Switzerland\\}

\begin{document}
\maketitle
\section{Introduction}
Large-scale survey telescope arrays have revolutionized astronomy, enabling high-precision, wide-scope observations \cite{Ofek_2023}. These systems collect massive volumes of data through collaborative telescope units, but their scale introduces challenges to system reliability and operational continuity. Robust monitoring and logging solutions are essential for handling high data throughput and enabling real-time insights \cite{Y.Zhong}. A leading example is the Cherenkov Telescope Array Observatory (CTAO) \cite{McMuldroch2024}, comprising over 100 telescopes across two sites—in Chile and Spain—controlled by the Array Control and Data Acquisition (ACADA) system \cite{Oya:2019byr}. ACADA oversees data handling and integrates the Science Alert Generation Pipeline (SAG) for early detection of transient event \cite{Panebianco2025}\cite{Castaldini2025}.

Ensuring the reliability of such systems is critical and extends beyond gamma-ray astronomy to other domains like radio and optical telescopes. To address this, we introduce ServiMon, a scalable, Docker-based \cite{docker} data collection pipeline designed for distributed environments like CTAO-ACADA. It leverages Prometheus \cite{prometheus}, Grafana \cite{grafana}, Cassandra \cite{carpenter2016cassandra}, and Kafka \cite{kafka} to provide real-time monitoring, interactive visualization, and fault detection. This paper presents ServiMon’s architecture and implementation, highlighting its role in enhancing system reliability, scalability, and laying the groundwork for predictive maintenance in astronomical infrastructures.

\section{Real-time Monitoring and Alerts: the CTAO Use Case}

The CTAO is an international project poised to revolutionize our understanding of the universe through the observation of very-high-energy (VHE) gamma rays. CTAO will enable scientists to explore cosmic phenomena with unprecedented precision, unveiling insights into the most energetic processes in the cosmos, such as supernovae, pulsars, black holes, and dark matter.

The CTAO is being developed by a global consortium of research institutions and agencies from more than 30 countries. The observatory consists of two arrays, strategically positioned in both the Northern hemisphere on La Palma Island, Spain, and the Southern hemisphere at Paranal, Chile. Together, these sites will host over 100 telescopes of three different sizes: Large-Sized Telescopes (LSTs), Medium-Sized Telescopes (MSTs), and Small-Sized Telescopes (SSTs), each optimized to cover different ranges of gamma-ray energies, from 20 GeV to over 300 TeV \cite{Vercellone2014CTA}.

From a technical standpoint, CTAO will use the imaging atmospheric Cherenkov technique, detecting the faint flashes of blue Cherenkov light produced when gamma rays interact with Earth's atmosphere. The high-speed cameras mounted on the telescopes will capture this transient light with nanosecond precision, generating a massive amount of data to be processed, filtered, and archived.

The scope of CTAO extends far beyond astrophysics; it is expected to contribute to fundamental physics and multi-messenger astronomy by complementing observations from gravitational wave detectors and neutrino observatories. 

CTAO is projected to produce up to several petabytes of data per year, depending on the observational conditions and the configuration of the array. This massive data volume requires a sophisticated and reliable data management system to ensure scientific integrity and operational efficiency.

At the heart of this challenge lies ACADA, the critical system responsible for orchestrating the operations of the entire observatory. ACADA handles the real-time control of telescopes, data acquisition, timing synchronization, health monitoring, and initial data filtering. Given the scale and complexity of CTAO, the full operability of ACADA is essential not only to coordinate the actions of the telescopes but also to ensure that no observational data are lost or corrupted during collection and transmission. Any malfunction or delay in this system could compromise irreplaceable scientific observations.

\section{ServiMon: Distributed Monitoring Architecture for Advanced System}
ServiMon integrates advanced tools to provide real-time monitoring and reporting in a distributed environment. Using state-of-the-art technologies like Prometheus, and Grafana, the system ensures modularity, scalability, and efficiency. This containerized cloud native architecture simplifies data collection, visualization, and management, enabling seamless oversight of complex systems.

\subsection{Architecture of the Monitoring Framework}
The monitoring system is built on a Docker-based infrastructure, providing isolation, portability, and ease of deployment across various environments. Each service operates in an independent container, minimizing resource interference and optimizing orchestration. The architecture integrates several key components that work together to ensure observability and performance monitoring across the system.

At the core of real-time metrics collection is \textbf{Prometheus}, which serves as the primary metrics aggregator. It gathers data from components such as Cassandra and Kafka using Java Management Extensions (JMX) Exporters \cite{jmx_exporter}. Through periodic scraping, Prometheus offers timely insights into both system and application-level performance.

For data visualization, \textbf{Grafana} plays a central role by offering a flexible and interactive interface. It connects to Prometheus for real-time metrics, enabling the creation of dynamic dashboards that can be adapted to various monitoring needs.

\subsection{Network Configuration for Secure Integration}
The architecture employs Docker networking to guarantee secure and isolated communication between components \cite{dockernetworking}. Custom configurations enable seamless internal communication while exposing selected ports for host-level access. Services like Prometheus, and Grafana are available through dedicated ports, ensuring secure data monitoring and logging.

\subsection{Scalable Architecture and Core Technologies}
The monitoring system is designed with scalability and fault tolerance at its core, making it well-suited for cloud-native deployments \cite{cloudnative}. It supports horizontal scaling and high availability, with key components such as Prometheus deployable in replicated configurations to sustain variable workloads and ensure reliable operation. Figure 1 outlines the system architecture, illustrating service interconnections and the role of exposed ports.

At the heart of the monitoring pipeline is \textbf{Prometheus}, which collects time-series data from components like \textbf{Cassandra} and \textbf{Kafka} via \textbf{JMX Exporters}. Prometheus supports robust querying through \textbf{PromQL} \cite{promql}, enabling fine-grained insights into system performance and resource usage.

Complementing this, \textbf{Grafana} delivers dynamic visualization capabilities, integrating both metrics and logs into customizable dashboards. It interfaces with \textbf{Prometheus} for metrics, providing a unified view for monitoring and troubleshooting.

The underlying services are also monitored in detail: \textbf{Cassandra}, a distributed NoSQL database, exposes metrics on latency, memory usage, and throughput via JMX, while \textbf{Kafka}, a distributed streaming platform, provides key performance indicators such as message throughput and broker health. These telemetry sources are essential for maintaining a comprehensive, real-time picture of system reliability and operational efficiency.

\begin{figure}[htbp]
\centering

\begin{subfigure}{\textwidth}
\centering
\includegraphics[height=0.3\textheight]{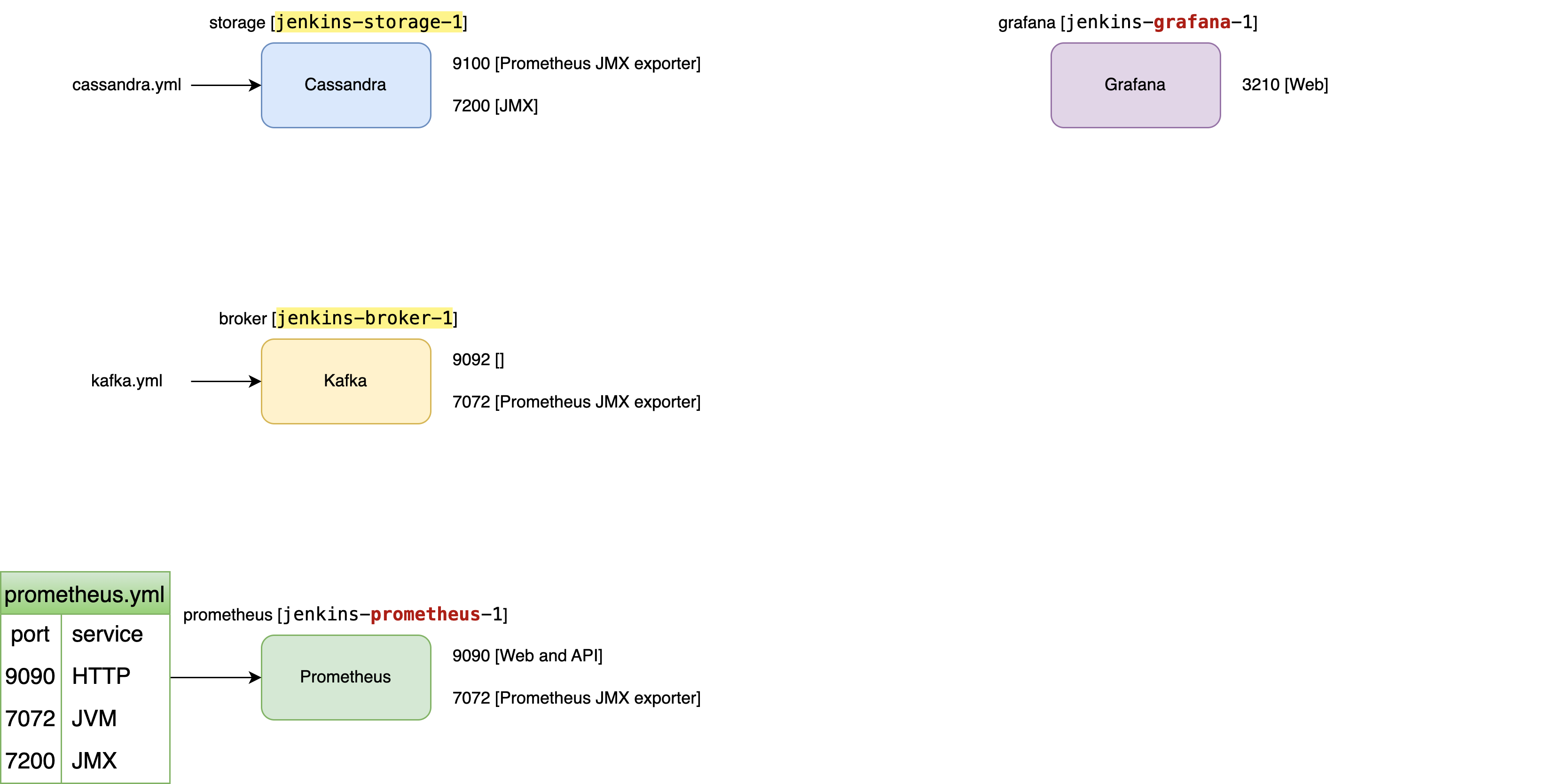}
\end{subfigure}
\caption{Monitoring architecture overview.}
\label{fig:servimon}
\end{figure}

\section{Implementation and Integration}
A series of Docker containers, each with a designated role in monitoring and logging, underpins the monitoring framework. The \textbf{service orchestration} is handled through Docker Compose, which streamlines the deployment and management of interdependent services. Its declarative configuration defines startup order and relationships among components such as Prometheus, Grafana, and JMX Exporters, ensuring proper sequencing and coordination. Docker Compose also enables scalability by allowing replica adjustments to accommodate varying loads and support high availability. For \textbf{container communication}, a Docker bridge network provides isolated and secure inter-connectivity. Each container is assigned a unique host-name to facilitate internal routing, while selected ports are exposed to the host for external access—for example, Prometheus on port 9090, and Grafana on 3210. This setup ensures reliable data exchange and enables Prometheus to access exporter metrics directly within the internal network. To guarantee \textbf{data persistence}, the system employs dedicated Docker volumes for each core component. Prometheus stores time-series metrics, and Grafana preserves dashboard configurations and user settings. These persistent volumes protect monitoring data and system state across container restarts, maintaining operational continuity even under dynamic conditions.

\section{Data Collection, Processing, and Integration}
The monitoring framework integrates real-time metrics collection, and data visualization to provide continuous insight into system performance and health. This cohesive approach ensures operational reliability across the distributed CTAO infrastructure.

Prometheus acts as the central time-series database, aggregating metrics from JMX Exporters linked to services like Cassandra and Kafka. These exporters expose detailed runtime metrics—such as latency, throughput, and resource usage—which Prometheus captures for longitudinal performance analysis.

Grafana provides the visualization layer, with interactive dashboards that integrate metrics from Prometheus. Users can monitor system status in real time and configure alerts to detect and respond to anomalies promptly.

The stack is integrated with CTAO systems via a custom Docker Compose setup, defining the data flow from instrumentation to visualization. Monitoring data from the array is streamed to Kafka for real-time processing and stored in Cassandra. Prometheus accesses data via JMX endpoints (ports 7072 and 7200), and Grafana then renders these data streams into clear, actionable visual outputs (Figure 2).

\begin{figure}[h]
\begin{subfigure}{0.5\textwidth}
\includegraphics[width=0.9\linewidth, height=6cm]{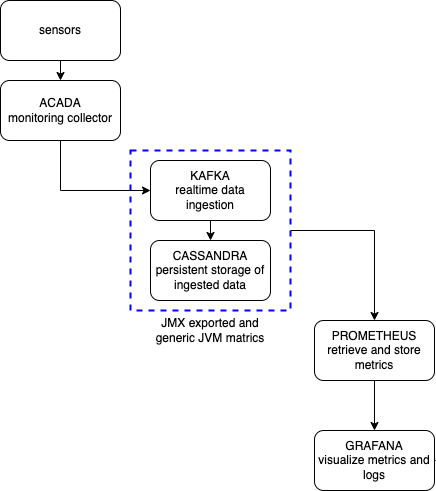} 
\caption{Information flow from data sources to visualization}
\label{fig:subim1}
\end{subfigure}
\begin{subfigure}{0.5\textwidth}
\includegraphics[width=0.9\linewidth, height=6cm]{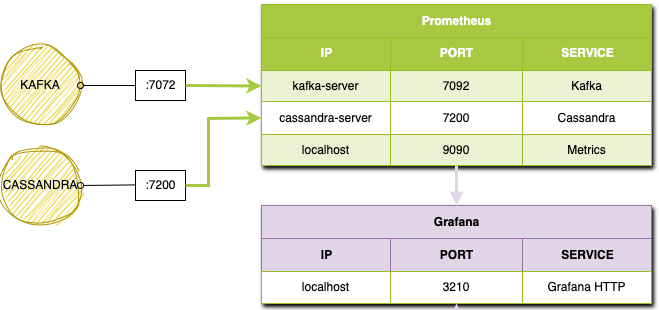}
\caption{Connections and data flow between monitoring tools}
\label{fig:subim2}
\end{subfigure}

\caption{Data Collection, Processing, and Integration}
\label{fig:image2}
\end{figure}

\subsection{Enhanced Monitoring with Grafana Alerting}
Grafana alerting can provide robust real-time anomaly and failure detection mechanisms. Users can create alert rules, which can then be visualized in Grafana panels and linked to alert conditions with defined thresholds and durations, ensuring timely and relevant notifications.

For large, complex deployments, alert configurations can be maintained in YAML files, enabling consistent rule management across environments and simplifying version control and automation workflows.

Simultaneously, Prometheus metrics are made accessible in Grafana by configuring Prometheus as a data source. Through PromQL, users can query specific performance indicators, apply filters, and define alert conditions, reinforcing system observability.

This dual-layered monitoring strategy—combining alerts and metrics—ensures robust fault detection and contributes to the operational reliability and productivity of the CTAO-ACADA Monitoring and Alarm subsystems.

\section{Predictive Maintenance}
The architecture of ServiMon establishes a solid foundation for implementing predictive maintenance \cite{predictivemaintenance} strategies within the dynamic context of the CTAO. By consolidating data from a broad range of sensors, logs, and events, ServiMon supports advanced forecasting tools essential for proactive system oversight.

The system’s data infrastructure opens the door to several predictive maintenance methodologies. Machine learning models, particularly recurrent neural networks (RNNs) and long short-term memory (LSTM) architectures, can be trained on historical data to predict failures by recognizing temporal patterns. In parallel, stochastic methods like hidden Markov models could provide probabilistic assessments of system health under varying conditions. 

By combining real-time metrics, and advanced predictive algorithms, the system offers a scalable, cost-effective framework for proactive maintenance. This approach can enhance the availability and longevity of critical components across the array, ensuring sustained scientific output and operational efficiency.

\section{Future Directions for ServiMon and Implications for Gamma-Ray Astronomy}
Although ServiMon’s predictive maintenance framework is still evolving, several promising directions for future development have been identified. One involves the integration of anomaly detection models directly into the monitoring pipeline, enabling real-time identification of system irregularities. Another enhancement lies in leveraging edge computing to pre-process data at its source, thereby reducing latency and network load while supporting faster, localized analytics. Additionally, refining machine learning algorithms through continuous operational feedback can significantly improve the precision and reliability of failure predictions.

These advancements will strengthen ServiMon’s scalability and resilience, extending its potential applications beyond CTAO to other large-scale scientific systems. At the core of these efforts is a robust data infrastructure, which remains essential for advancing predictive strategies in complex technical environments.

Implementing predictive maintenance within ServiMon will be instrumental in maximizing the uptime and efficiency of the CTAO. This reliability is critical for maintaining its role as a leading observatory in gamma-ray astronomy. By reducing operational disruptions and enabling consistent data collection, ServiMon will contribute directly to the scientific output and long-term success of the array, supporting future discoveries in high-energy astrophysics.

\section{Conclusion}
ServiMon is a modular, service-oriented monitoring framework designed to meet the complex demands of distributed infrastructures like the CTAO. By integrating technologies such as Kafka, Cassandra, Prometheus, and Grafana, it ensures efficient data streaming and intuitive visualization, enabling real-time performance monitoring and fault detection.

The combination of real-time telemetry with historical data facilitates advanced strategies, such as anomaly detection and early warnings, which are crucial for minimizing downtime and optimizing maintenance. Its scalable architecture allows ServiMon to adapt to various infrastructures or telescopes, rather than just the CTAO.

Future development will concentrate on enhancing predictive maintenance through deep learning models like LSTMs and GRUs, as well as hybrid approaches that merge machine learning with stochastic methods.

Thanks to its adaptability, ServiMon is well-positioned for broader application in astronomical observatories. Its ongoing evolution will help ensure the reliability, efficiency, and scientific output of next-generation gamma-ray facilities.

\section{Acknowledgments}
We gratefully acknowledge financial support from the agencies and organizations listed here:

https://www.ctao.org/for-scientists/library/acknowledgments/

Additional support was provided by the Italian Research Center on High Performance Computing, Big Data, and Quantum Computing (ICSC), project funded by European Union - NextGenerationEU - and National Recovery and Resilience Plan (NRRP) - Mission 4 Component 2 within the activities of Spoke 2 (Fundamental Research and Space Economy).

\bibliography{servimon_icrc}

\begin{thebibliography}{10}

\bibitem{Ofek_2023}
Eran~O. Ofek et~al.
\newblock {The Large Array Survey Telescope—System Overview and Performances}.
\newblock {\em Publications of the Astronomical Society of the Pacific}, 2023.

\bibitem{Y.Zhong}
Y.~Zhong, G.~Lyu, X.~He, Y.~Zhang, and S.~S. Ge.
\newblock {Distributed Active Fault-Tolerant Cooperative Control for Multiagent Systems With Communication Delays and External Disturbances}.
\newblock {\em IEEE Transactions on Cybernetics}, 53(7):4642--4652, July 2023.

\bibitem{McMuldroch2024}
S.~McMuldroch et~al.
\newblock {The Cherenkov Telescope Array Observatory (CTAO): ongoing development of the world’s premier ground-based gamma ray observatory}.
\newblock In {\em Proceedings of SPIE}, volume 13094, page~19, 2024.

\bibitem{Oya:2019byr}
Igor Oya et~al.
\newblock {The Array Control and Data Acquisition System of the Cherenkov Telescope Array}.
\newblock In {\em {17th International Conference on Accelerator and Large Experimental Physics Control Systems}}, page WEMPR005, 2020.

\bibitem{Panebianco2025}
G.~Panebianco et~al.
\newblock {SAG-SCI: the Real-time, High-level Analysis Software for Array Control and Data Acquisition of the Cherenkov Telescope Array Observatory}.
\newblock In {\em Proc. ICRC 2025}.

\bibitem{Castaldini2025}
L.~Castaldini et~al.
\newblock {The new architecture design of the Science Alert Generation of the Cherenkov Telescope Array Observatory}.
\newblock In {\em Proc. ICRC 2025}.

\bibitem{docker}
Dirk Merkel.
\newblock {Docker: Lightweight Linux Containers for Consistent Development and Deployment}.
\newblock {\em Linux Journal}, 2014(239):2, 2014.

\bibitem{prometheus}
Bjorn Rabenstein and Julius Volz.
\newblock {Prometheus: A Next-Generation Monitoring System (Talk)}.
\newblock In {\em USENIX Association}, 2015.

\bibitem{grafana}
Grafana Labs.
\newblock {\em Grafana Documentation}, July 2019.

\bibitem{carpenter2016cassandra}
J.~Carpenter and E.~Hewitt.
\newblock {\em Cassandra: the Definitive Guide}.
\newblock O'Reilly, 2016.

\bibitem{kafka}
R.~Shree, T.~Choudhury, S.~C. Gupta, and P.~Kumar.
\newblock {Kafka: The Modern Platform for Data Management and Analysis in Big Data Domain}.
\newblock In {\em 2017 2nd International Conference on Telecommunication and Networks (TEL-NET)}, pages 1--5, Noida, India, 2017.

\bibitem{Vercellone2014CTA}
Stefano Vercellone.
\newblock {The next generation Cherenkov Telescope Array observatory: CTA}.
\newblock {\em arXiv preprint}, arXiv:1405.5696, 2014.
\newblock Covers Large-, Medium- and Small-Sized Telescopes (LST, MST, SST).

\bibitem{jmx_exporter}
Prometheus JMX~Exporter Maintainers.
\newblock {JMX Exporter: A collector to capture JMX MBean metrics for Prometheus}.
\newblock GitHub repository, 2025.
\newblock Available at GitHub.

\bibitem{dockernetworking}
Docker Documentation.
\newblock Networking overview, 2024.

\bibitem{cloudnative}
Y.~Mao, Y.~Fu, S.~Gu, S.~Vhaduri, L.~Cheng, and Q.~Liu.
\newblock {Resource Management Schemes for Cloud-Native Platforms with Computing Containers of Docker and Kubernetes}.
\newblock arXiv preprint arXiv:cs.DC/2020, 2020.

\bibitem{promql}
Prometheus.
\newblock Querying basics | prometheus, 2024.

\bibitem{predictivemaintenance}
M.~Patel, J.~Vasa, and B.~Patel.
\newblock {Predictive Maintenance: A Comprehensive Analysis and Future Outlook}.
\newblock In {\em 2023 2nd International Conference on Futuristic Technologies (INCOFT)}, pages 1--7, Belagavi, Karnataka, India, 2023.

\end{thebibliography}

\end{document}